\documentclass[aoas,MSNbibl,nameyear,seceqn,dvips]{arximspdf}
\usepackage{multirow,dcolumn,url,breakurl}
\usepackage{graphicx}

\doi{10.1214/14-AOAS736} 
\volume{8}
\issue{2}
\pubyear{2014}
\firstpage{1045}
\lastpage{1064}

\makeatletter
\newcolumntype{d}[1]{D{.}{.}{#1}}
\newcommand{\widebar}{\overline}
\makeatother

\begin{document}
\begin{frontmatter}

\title{Regularized 3D functional regression for brain image data via
Haar wavelets\thanksref{T11}}
\runtitle{Regularized functional regression}

\begin{aug}
\author[a]{\fnms{Xuejing} \snm{Wang}\thanksref{t1}\ead[label=e1]{xuejwang@umich.edu}},
\author[a]{\fnms{Bin} \snm{Nan}\corref{}\thanksref{t1,t2}\ead[label=e2]{bnan@umich.edu}},
\author[b]{\fnms{Ji} \snm{Zhu}\thanksref{t1,t3}\ead[label=e3]{jizhu@umich.edu}},
\author[c]{\fnms{Robert} \snm{Koeppe}\thanksref{t1}\ead[label=e4]{koeppe@umich.edu}}\\
\and
\author{\snm{for the Alzheimer's Disease Neuroimaging Initiative}}
\runauthor{X. Wang et al.}
\affiliation{University of Michigan}
\address[a]{X. Wang\\
B. Nan\\
Department of Biostatistics\\
University of Michigan\\
1415 Washington Heights\\
Ann Arbor, Michigan 48109-2029\\
USA\\
\printead{e1}\\
\phantom{E-mail:\ }\printead*{e2}}
\address[b]{J. Zhu\\
Department of Statistics\\
University of Michigan\\
1085 South University Ave \\
Ann Arbor, Michigan 48109-1107\\
USA\\
\printead{e3}}
\address[c]{R. Koeppe\\
Department of Radiology\\
University of Michigan\\
1500 E. Medical Center Drive\\
Ann Arbor, Michigan 48109-5030\\
USA\\
\printead{e4}} 
\end{aug}
\thankstext{T11}{Supported by NIH Grants P30 AG010129 and K01 AG030514.}
\thankstext{t1}{Supported in part by NIH Grant R01-AG036802.}
\thankstext{t2}{Supported in part by NSF Grant DMS-1007590.}
\thankstext{t3}{Supported in part by NSF Grant DMS-0748389.}

\received{\smonth{7} \syear{2012}}
\revised{\smonth{2} \syear{2014}}

%
\begin{abstract}
The primary motivation and application in this article come from brain
imaging studies on
cognitive impairment in elderly subjects with brain disorders. We
propose a regularized
Haar wavelet-based approach for the analysis of three-dimensional brain
image data in the
framework of functional data analysis, which automatically takes into
account the spatial
information among neighboring voxels. We conduct extensive simulation
studies to evaluate
the prediction performance of the proposed approach and its ability to
identify related
regions to the outcome of interest, with the underlying assumption that
only few relatively
small subregions are truly predictive of the outcome of interest. We
then apply the proposed
approach to searching for brain subregions that are associated with
cognition using PET images
of patients with Alzheimer's disease, patients with mild cognitive
impairment and normal controls.
\end{abstract}

%

\end{frontmatter}

\section{Introduction}\label{sec1}
Alzheimer's disease (AD) is the most frequent cause of dementia in our
increasingly aging societies, representing a significant impact on the
US population with 10\% prevalence in individuals aged above 70 years old
[\citet{plassman07}]. Despite the prevalence, this disease remains
quite a mystery; there is neither a cure nor a definite treatment to
arrest its course and, currently, the only definite way to diagnose AD
is to examine the brain tissue after death. According to recent studies
[\citet{Leifer03}], early diagnosis of AD is of great value since
new drug therapies can be used to potentially delay the progression of
the disease. To this end, much progress has been made in assisting the
early diagnosis of AD with neuroimaging techniques. One such widely
used neuroimaging technique is positron emission tomography (PET)
imaging, which is one of the most promising tools for the early
diagnosis of AD, and
it is of great scientific interest in understanding the association
between PET images and cognitive impairment.
In particular, the fluorodeoxyglucose (FDG) PET has been used to
measure the cerebral glucose metabolic activity for over 20 years.

FDG PET scans used in the preparation of this article were obtained
from a large multi-center follow-up study on Alzheimer's disease and
early dementia, the Alzheimer's Disease Neuroimaing Initiative (ADNI).
A total of 403 FDG PET scans were acquired for this application,
including 102 normal control (NC) subjects, 206 subjects with mild
cognitive impairment (MCI) and 95 subjects diagnosed with AD. In this
study, we consider the baseline FDG PET scans with a standard $160\times
160\times96$ voxel image grid as the predictor to the cognitive
performance as measured by the mini-mental state exam (MMSE), which is
a questionnaire test that is used to screen for cognitive impairment
[\citet{Cockrelletal88}]. The maximum MMSE score is 30 and, on
average, MMSE scores decline as the disease progresses. The goal of our
study is to identify brain subregions that are most closely related to
the prediction of MMSE scores.

Many methods have been developed for the analysis of brain image data
in order to identify disease-related brain subregions. Most of these
methods focus on region of interest (ROI) and voxel-based univariate
analysis; see, for example, \citet{Luoetal03}, \citet
{Grimmeretal09} and \citet{Shinetal10}, among many others. For AD
in particular, several studies have shown that reduced metabolic
activity in some regions of the brain, such as the posterior cingulate
and the temporal and parietal cortices, are associated with the
progression of cognitive impairment [\citet{Fosteretal84},
\citeauthor{Minoshimaetal95} (\citeyear{Minoshimaetal95,Minoshimaetal97})]. These methods are intended to
provide statistics by doing a separate analysis for each ROI or voxel
and then to draw inferences at the region- or voxel-level.
As a result of testing millions of hypotheses,
appropriate adjustments for multiple comparisons have to be considered.
In the neuroimaging literature, a distinction is often drawn between
such univariat analyses and an alternative, multiple covariate
regression models that treat every voxel as a covariate. Since the
number of voxels is much larger than the number of scans, the ordinary
least squares for linear regression cannot be implemented without
applying, for example, some dimension reduction techniques. Such
analysis, however, may lead to difficulties in interpretations and
practical implications. Both the traditional univariate and multiple
covariate approaches (if applicable) have one major limitation in
common: they are developed without considering the spatial information
of the brain, possibly resulting in some loss of information. There is
an emerging awareness of the importance of taking such information into
account. For example, multiple covariate analysis can be conducted with
a focus on extracting principal components from the images [\citet
{Fetal96}, \citet{Kerroucheetal06}]. More recently, a variety of
Bayesian spatial modeling approaches have been proposed to model the
correlation between neighboring voxels, which need to carefully specify
the prior distributions; see, for example, \citet{Bowmanetal08},
\citet{Kangetal11}. Our way of addressing this issue is to
treat the entire 3D image as a single functional input, which allows
retaining all the information from the original image without modeling
the spatial correlation between voxels explicitly.
Specifically, in this article, we treat PET image data as the 3D
functional observations, and propose a novel Haar wavelet-based
regularized approach to analyze PET image data in the framework of
functional data analysis.

Functional regression models are known as one of the standard
techniques in functional data analysis. It is noted that the models can
be defined as functional in one or both of two ways: the response
variable is functional; at least one of the covariates is functional.
In this article, we focus on the functional linear regression model
with a scalar response variable and a single functional predictor.
Using the 1D case as an illustration, the functional linear regression
model relates a scalar response variable $Y$ to a functional predictor
$X(t)$ as follows:
%
\begin{equation}
\label{1dmodel} Y_{i} =\beta_{0}+ \int_{0}^{T}
X_{i}(t)\beta(t) \,dt+\varepsilon_{i}, \qquad i=1,\ldots,n,
\end{equation}
where $\beta(t)$ is the regression coefficient function and $t$ refers
to time or location. For the 3D case we consider later, $t$ is replaced
by the coordinate $(u,v,w)$. Regularization methods, such as the
roughness penalty approach or using restricted basis functions
[\citet{ram97}], can be implemented to produce an estimator of
$\beta(t)$ that is meaningful in interpretation and useful in prediction.

For the functional linear regression model (\ref{1dmodel}), \citet
{Jamesetal2009} proposed a regularized approach that focuses on
producing sparse and highly interpretable estimates of the coefficient
function $\beta(t)$. This approach involves first dividing the domain
into a fine grid of points, and then using appropriate variable
selection methods to determine whether the $d$th derivative of $\beta
(t)$ is zero or not at each of the grid points, that is, $\beta
^{(d)}(t)=0$ for one or more values of $d \in\{0,1,2,\ldots\}$. They
proposed the Dantzig selector [\citet{Candes07}] and a Lasso-type
approach for the estimation of $\beta(t)$ using piecewise constant
basis, where the Dantzig selector seems to be more natural. Empirical
results show that their methods perform well when $p$, the number of
basis functions, is not too large.
When functional data are measured over a very fine grid such as brain
image data, the Dantzig selector faces the challenge of solving a huge
linear programming problem and the Lasso-type algorithm can be\vadjust{\goodbreak}
extremely slow; note that for the latter the fast coordinate descent
algorithm [\citet{Fu98}, \citet{Daub04}, \citet
{Friedmanetal07}, \citet{Wu08}] does not apply due to the penalty
on derivatives. Without imposing sparsity, \citet{ReissOgden2010}
considered the functional principal component regression for image
data.\looseness=-1

In this article, we choose the Haar wavelet basis instead of the
piecewise constant basis for analyzing 3D image data and show that the
Haar wavelet-based approach presents a number of advantages. First, it
yields regional sparseness without imposing constraints on derivatives,
which is needed in \citet{Jamesetal2009}. In other words, by
shrinking corresponding wavelet coefficients to zero, the estimator of
the regression coefficient function can be exactly zero over regions
where no relationship to the response variable is present.
Second, the Haar wavelet transform offers a way to overcome the issue
of high multicollinearity caused by high neighboring spatial correlations.
Third, our approach is flexible enough to allow the coefficient
function to be estimated at different levels of smoothness through
choosing different levels of the Haar wavelet decomposition. Fourth,
the Haar wavelet transform can be applied as a dimension reduction
technique prior to model fitting for high-dimensional image data by
setting a common set of close to zero wavelet coefficients of PET
images to zero, which is an effective way of removing voxels outside
the brain or in the ventricles. It should be noted that a recent
article by \citet{zhaoetal12} considered a general wavelet-based
Lasso approach in functional linear regression, but only concerned 1D
$\beta(t)$. The Haar wavelet transform is a useful tool for image and
signal analysis and has many other applications. For example,
\citet{Sauxetal12} and \citet{Lovejoyetal12} discussed the
use of Haar wavelet transforms in geophysics and climate research.

The rest of this article is organized as follows. In Section~\ref{sec2} we
review some background on wavelet decomposition and properties of Haar
wavelet basis functions using a 1D functional linear regression model
as an illustration and then propose the $\ell_1$ regularized shrinkage
estimation for general functional data, including both 1D and 3D cases.
To evaluate the numerical performance of our approach, we conduct
extensive simulations in Section~\ref{sec3}.
We present the analysis of ADNI 3D PET image data in Section~\ref{sec4} and make
some concluding remarks in Section~\ref{sec5}. We also show that the proposed
method achieves the desirable nonasymptotic error bounds for prediction
and estimation, the so-called oracle inequalities, meaning that the
method performs equally well (up to a constant) as if the true
subregions with nonzero regression coefficient were given. The
theoretical results are provided in the online supplementary material
[\citet{Wangetal2014}].

\section{Regularized Haar wavelets method}\label{sec2}
For ease of presentation, we describe the proposed methodology starting
with the 1D case given in (\ref{1dmodel}), then extend it to the 3D
case using a tensor product of three 1D wavelet expansions.

\subsection{Choice of basis}\label{sec2.1}

Basis expansions are commonly used in analyzing functional data.
Among a variety of choices of basis expansions, wavelets have the
important ability to allow simultaneous time (or space in this article)
and frequency localization. Unlike many other commonly used basis
systems, wavelet transforms are highly adaptable to different levels of
smoothness and more capable of capturing edges, spikes and other types
of discontinuities, especially for wavelet transforms with relatively
small support such as the Haar wavelets. Wavelet transforms also
provide a powerful tool to compress the data. A compressed
approximation of the signal can be achieved by penalizing the wavelet
coefficients [\citet{Wandetal11}], which involves shrinking small
coefficients to zero and possibly shrinking the large ones without
affecting the main features of the data. Hence, it is advantageous to
use wavelet transforms to decompose images as well as the regression
coefficient function for estimation.

In many applications, it is often the case that the association between
$X(t)$ and $Y$ in model (\ref{1dmodel}) is sparse and potentially
discontinuous at the boundaries of subregions. In particular, only few
brain subregions in the aforementioned PET images are believed to be
related to cognitive impairment.
To better identify such patterns, we choose to use Haar wavelets. The
Haar wavelet transform is easily calculated and affected less by
discontinuities. In addition, sparsity of $\beta(t)$ can be recovered
by shrinking its wavelet coefficients to zero.
The scaling function (also called a father wavelet) $\phi$ and the
mother wavelet $\psi$ of Haar wavelets defined on $[0,1)$ are given below:
\begin{eqnarray*}
\phi(t) &=& \cases{ 1, &\quad if $0 \leq t < 1$;
\vspace*{3pt}\cr
0, &\quad otherwise;}
\\
\psi(t) &=& \cases{ 1, &\quad if $0 \leq t <1/2$;
\vspace*{3pt}\cr
-1, &\quad if $1/2 \leq t
<1$;
\vspace*{3pt}\cr
0, &\quad otherwise.}
\end{eqnarray*}
The Haar wavelet bases are then generated in the form of translations
and dilations of the above father and mother wavelet functions as
\begin{eqnarray*}
\phi_{j,k}(t)&=&\sqrt{2^{j}}\phi\bigl(2^{j}t-k
\bigr),
\\
\psi_{j,k}(t)&=&\sqrt{2^{j}}\psi\bigl(2^{j}t-k
\bigr),
\end{eqnarray*}
where $j=0,1,\ldots$ and $k=0,1, \ldots,2^{j}-1$. The index $j$ refers
to dilations and $k$ refers to translations and $\sqrt{2^j}$ is the
normalizing factor. It is noted that the basis functions are
orthonormal. Therefore, for a sufficiently fine resolution $J$, the
coefficient function $\beta(t)$ in (\ref{1dmodel}) defined on $[0, 1)$
can be expanded in a Haar wavelet series:
%
\begin{equation}
\label{beta-wavelet} \beta(t)=\sum_{k=0}^{2^{j_{0}}-1}
a_{j_{0},k}\phi_{j_{0},k}(t)+ \sum_{j=j_{0}}^{J}
\sum_{k=0}^{2^{j}-1}d_{j,k}
\psi_{j,k}(t)+e(t),
\end{equation}
where\vspace*{1pt} $a_{j_{0},k}=\int_{0}^{1} \beta(t)\phi_{j_{0},k}(t) \,dt$ are the
approximation coefficients at the coarsest resolution $j_{0}$,
$d_{j,k}=\int_{0}^{1} \beta(t)\psi_{j,k}(t) \,dt$ are the detail
coefficients that characterize the finer structures of $\beta(t)$ as
$j$ grows, and $e(t)$ is the approximation error that goes to zero as
$J$ goes to infinity. The Haar wavelet representation of a signal thus
consists of approximations together with details that can provide the
desirable frequencies. See, for example, \citet{Walker08} for more
details about Haar wavelets.

\subsection{Model estimation}\label{sec2.2}
Rewrite $\beta(t)$ in (\ref{beta-wavelet}) by
%
\begin{equation}
\label{beta} \beta(t)=B(t)^T \eta+e(t),
\end{equation}
where $B(t)$ denotes the collection of all $\phi_{j,k}(t)$ and $\psi
_{j,k}(t)$ in the above Haar wavelet expansion, and $\eta$ is
the corresponding wavelet coefficient vector of length~$p$.
Plugging (\ref{beta}) into (\ref{1dmodel}), we obtain
%
\begin{equation}
\label{linear-model}
\qquad\quad Y_{i} =\beta_{0}+ \int_{0}^{1}
X_{i}(t)B(t)^T \eta \,dt+\varepsilon_{i}^{*}=
\beta_{0}+C_{i}^T\eta+\varepsilon_{i}^{*},
\qquad i=1,\ldots,n,
\end{equation}
where $C_{i}=\int_{0}^{1}X_{i}(t)B(t) \,dt $ and $\varepsilon_{i}^{*} =
\varepsilon_i + \int_0^1 X_i(t)e(t) \,dt$. It should be noted that $C_{i}$
is the wavelet coefficient vector of $X_i(t)$ when we decompose
$X_{i}(t)$ using the same set of Haar wavelet basis functions as those
in (\ref{beta}). Model (\ref{linear-model}) can then be rewritten as follows:
%
\begin{equation}
\label{model} Y=\beta_{0}+C\eta+\varepsilon^{*},
\end{equation}
where $C={}[ C_{1},C_{2},\ldots,C_{n}] ^T$ is an ${n\times p}$ design
matrix in linear model (\ref{model}). Once an estimator $\hat\eta$ is
obtained from (\ref{model}), $\beta(t)$ can then be estimated by
$B(t)^T\hat\eta$.

In practice, $X(t)$ is observed on only a finite set of grid points $\{
t_{1},\ldots,t_{p}\}$, which also determines the highest and yet
practically meaningful level of decomposition for $\beta(t)$. For the
discrete wavelet transform, $p$ is required to be a power of 2.
Using the usual terminology for Haar wavelets [see, e.g., \citet
{Walker08} and that used in the MATLAB Wavelet Toolbox (2011b)], we
define the level 1 Haar wavelet decomposition by computing the average
and the difference on each consecutive pair of values, and the maximum
level is $\log_{2}p$. The level number is directly determined by the
integer $j_0$ in (\ref{beta-wavelet}). For any level of Haar wavelet
decomposition, the total number of basis functions $\phi_{j,k}$ and
$\psi_{j,k}$ is always $p$, and the collection of $\phi_{j,k}$ and $\psi
_{j,k}$ then forms a set of $p$-dimensional orthonormal basis functions.

A key advantage of using Haar wavelets is as follows. When $\beta(t)=0$
in large regions of $t\in[0,1)$ (in the ADNI brain image analysis
where $t$ is 3D, this would correspond to that large regions in the
brain are not associated with the cognitive performance measured by
MMSE), the coefficient vector $\eta$ in (\ref{beta}) should be sparse
with $e(t)=0$ for those regions, that is, $\beta(t)$ can be well
approximated by an economical wavelet expansion with few nonzero coefficients.
We consider the Lasso approach [\citet{tib96}] and implement the
method with the fast coordinate descent algorithm to obtain a desirable
sparse solution for the wavelet coefficients.

For a given $j_0$, which corresponds to a specific level of Haar
wavelet expansion, the Lasso estimator for $\eta$ is given by
%
\begin{equation}
\label{lasso} \hat{\eta}=\mathop{\arg\min}_{\eta} \biggl\{
\frac{1}{n}\|Y-\beta_{0}-C\eta\|_{2}^{2}+2
\lambda\|\eta\|_{1} \biggr\},
\end{equation}
where $\|\cdot\|_{1}$ and $\|\cdot\|_{2}$ denote the $\ell_{1}$ and
$\ell_{2}$ norms, respectively, and $\lambda\geq0$ is a tuning
parameter. In our estimating procedure, $j_0$ is also a tuning parameter.

It should be noted that in general the Haar wavelet coefficients with
large magnitudes are related to salient features. The magnitudes of
detail coefficients should be proportional to the differences between
every pair of values, that is, larger magnitudes indicate sharper
changes at corresponding locations and zero magnitudes indicate no
change. If both detail and approximation coefficients of the Haar
wavelet transform are close to zero, then $\beta(t)$ is close to zero.
Thus, we are able to obtain a sparse solution of $\beta(t)$ by
shrinking its small wavelet coefficients to zero.

\subsection{Selection of tuning parameters}\label{sec2.3}
In addition to the Lasso tuning parameter $\lambda$ in (\ref{lasso}),
we also need to take into account the level of the Haar wavelet
decomposition. There should exist an optimal level of decomposition for
$\beta(t)$ in terms of certain criterion, such as AIC, BIC or
cross-validation. If the length of observed $X_{i}(t)$ is $p$, then the
maximum possible level of the discrete Haar wavelet transform is $\log
_{2}p$, which is relatively small. Moreover, lower levels are usually
considered in real applications. Therefore, including two tuning
parameters does not increase computational burden by much.

\subsection{3D case}\label{sec2.4}

The ADNI's FDG PET brain images are 3D. A 3D function can be decomposed
using a tensor product of three 1D Haar wavelets. In particular, the 3D
Haar wavelet transform can be considered as averaging and differencing
operations [\citet{Muraki92}]. The averaging operation is
constructed by the 3D scaling function below:
\[
\label{3dphi} \phi_{j,\{k,l,m\}}(u,v,w)=\phi_{j,k}(u)
\phi_{j,l}(v)\phi_{j,m}(w).
\]
The differencing operation is taken in seven directions constructed by
the 3D wavelet functions as follows:
\begin{eqnarray*}
\label{3dpsi} \psi_{j,\{k,l,m\}}^{1}(u,v,w)&=&\phi_{j,k}(u)
\phi_{j,l}(v)\psi_{j,m}(w),
\nonumber
\\
\psi_{j,\{k,l,m\}}^{2}(u,v,w)&=&\phi_{j,k}(u)
\psi_{j,l}(v)\phi_{j,m}(w),
\nonumber
\\
\psi_{j,\{k,l,m\}}^{3}(u,v,w)&=&\phi_{j,k}(u)
\psi_{j,l}(v)\psi_{j,m}(w),
\nonumber
\\
\psi_{j,\{k,l,m\}}^{4}(u,v,w)&=&\psi_{j,k}(u)
\phi_{j,l}(v)\phi_{j,m}(w),
\nonumber
\\
\psi_{j,\{k,l,m\}}^{5}(u,v,w)&=&\psi_{j,k}(u)
\phi_{j,l}(v)\psi_{j,m}(w),
\nonumber
\\
\psi_{j,\{k,l,m\}}^{6}(u,v,w)&=&\psi_{j,k}(u)
\psi_{j,l}(v)\phi_{j,m}(w),
\nonumber
\\
\psi_{j,\{k,l,m\}}^{7}(u,v,w)&=&\psi_{j,k}(u)
\psi_{j,l}(v)\psi_{j,m}(w).
\end{eqnarray*}

Let the image $X_{i}(u,v,w)$ be a 3D functional predictor and $Y_{i}$
be a scalar response variable (MMSE, e.g.) for subject $i$,
$i=1,\dots, n$. The 3D functional linear regression model can be
written as
%
\begin{equation}
\label{3dmodel} Y_{i} =\beta_{0}+ \int_{0}^{T_{1}}\!
\int_{0}^{T_{2}}\!\int_{0}^{T_{3}}
X_{i}(u,v,w)\beta(u,v,w) \,du\,dv\,dw+\varepsilon_{i}.
\end{equation}
For a sufficiently fine resolution $J$, the 3D coefficient function
$\beta(u,v,w)$ can be approximated by
%
\begin{eqnarray}
\label{3dbeta} && \sum_{k,l,m=0}^{2^{j_{0}}-1}
a_{j_{0},\{k,l,m\}}\phi_{j_{0},\{k,l,m\}}(u,v,w)
\nonumber\\[-8pt]\\[-8pt]
&&\qquad{}+ \sum_{j=j_{0}}^{J}\sum
_{k,l,m=0}^{2^{j}-1}\sum
_{q=1}^{7}d_{j,\{k,l,m\}}^{q}
\psi_{j,\{k,l,m\}}^{q}(u,v,w).
\nonumber
\end{eqnarray}
Denote the set of all basis functions $\phi_{j,\{k,l,m\}}$ and $\psi
_{j,\{k,l,m\}}^{q}$ in (\ref{3dbeta}) by $B(u,v,w)$ and the wavelet
coefficients in (\ref{3dbeta}) by $\eta$, then $\beta(u,v,w)$ can be
written as
%
\begin{equation}
\label{3dbetasim} \beta(u,v,w)=B(u,v,w)^{T}\eta+e(u,v,w).
\end{equation}
Plugging (\ref{3dbetasim}) into model (\ref{3dmodel}), we obtain
%
\begin{eqnarray}
\label{3dmodelsim} Y_{i} &=& \beta_{0}+ \int
_{0}^{T_{1}}\!\int_{0}^{T_{2}}\!
\int_{0}^{T_{3}} X_{i}(u,v,w)B(u,v,w)^{T}
\eta \,du\,dv\,dw+\varepsilon_{i}^{*}
\nonumber\\[-8pt]\\[-8pt]
&=& \beta_{0}+C_{i}^T\eta+
\varepsilon_{i}^{*},
\nonumber
\end{eqnarray}
where $C_{i}=\int_{0}^{T_{1}}\!\int_{0}^{T_{2}}\!\int_{0}^{T_{3}}
X_{i}(u,v,w)B(u,v,w) \,du\,dv\,dw$, which is equivalent to the wavelet
coefficient vector when we apply the 3D wavelet transform to
$X_{i}(u,v,w)$, and $\varepsilon_i^* = \varepsilon_i + \int_{0}^{T_{1}}\!\int
_{0}^{T_{2}}\!\int_{0}^{T_{3}} X_{i}(u,v,w)e(u,v,w) \,du\,dv\,dw$. Then the
methodology proposed in the previous subsections for the 1D case can be
applied directly.

Following the calculations of \citet{Bickeletal09}, we can show
that our proposed method also enjoys the nonasymptotic oracle
inequalities similar to the linear model with high-dimensional
covariates. All the theoretical results are provided in the online
supplementary material [\citet{Wangetal2014}].\vadjust{\goodbreak} It should be noted
that the results are derived using the 1D notation for the estimator
$\hat{\beta}(t)$ for simplicity, but they hold exactly for the 3D case
of $\hat{\beta}(u,v,w)$.

\section{Simulation studies}\label{sec3}
To investigate the performance of the proposed Haar wavelet-based
approach, we have conducted extensive simulations for both 1D~and~3D
functional data. The results for 1D cases can be easily visualized,
whereas the 3D case mimics the brain images more naturally.

\subsection{1D simulation}\label{sec3.1}
We consider a variety of settings of $X(t)$ and $\beta(t)$. For $X(t) =
X^{*}(t) + {\mathcal E}(t)$ defined on $0\leq t \leq1$, where
${\mathcal E}(t)\sim N(0,\sigma_{\mathcal E}^2)$ is the noise term
independent of time $t$, we consider the following two scenarios:
\begin{itemize}
\item[--] Fourier: $X^{*}(t)=a_{0}+a_{1}\sin(2\pi t)+a_{2}\cos(2\pi
t)+a_{3}\sin(4\pi t)+a_{4}\cos(4\pi t)$.

\item[--] $B$-splines: $X^{*}(t)$ is a linear combination of cubic
$B$-splines with interior knots at $1/7,\ldots, 6/7$ and coefficients
$a_i$, that is, $X^*(t)=\sum a_{i}\phi_{i}(t)$, where $\phi_{i}(t)$ are
the $B$-spline basis functions.
\end{itemize}
In both scenarios, the coefficients $a_{i} \sim N(0,1)$. To assess the
performance of the proposed approach in identifying continuous and
discontinuous signals, we consider two cases of the regression
coefficient function $\beta(t)$:
\begin{itemize}
\item[--] \textit{Case} 1: $\beta(t)$ is a smooth function,
\[
\beta(t) = \cases{ 0.5 \bigl(\sin(20t-\pi)+1 \bigr), &\quad if $\pi/8
\leq t <
9 \pi/40$,
\vspace*{3pt}\cr
0, &\quad otherwise.}
\]

\item[--] \textit{Case} 2: $\beta(t)$ is piecewise constant,
\[
\beta(t) = \cases{ 1, &\quad if $0.2 \leq t < 0.3$,
\vspace*{3pt}\cr
0.5, &\quad if $0.5
\leq t < 0.7$,
\vspace*{3pt}\cr
0, &\quad otherwise.}
\]
\end{itemize}

\begin{figure}

\includegraphics{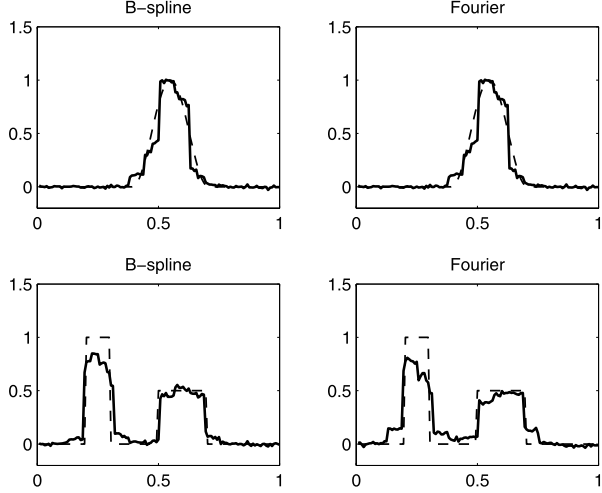}

\caption{Average of $\hat{\beta}(t)$ estimated using 5-fold
cross-validation with 100 replications (solid line). The dashed line is
the true $\beta(t)$. The top panel is for case~1, and the bottom panel
is for case~2.}\label{fig1}
\end{figure}

For each curve $X^{*}(t)$, we record $p=128$ equally spaced
measurements for convenience. The variance of the noise term ${\mathcal
E}(t)$ is set to be $\sigma_{\mathcal E}^2 = \frac{1}{p-1}\sum
_{j=1}^{p}(X^{*}(t_{j})-\widebar{X}^{*}(t_j))^2$, where $\widebar{X}^{*}(t_j)$
is the mean of $X^{*}(t_{j})$. The error\vspace*{1pt} term $\varepsilon$ in model (\ref
{1dmodel}) also follows a normal distribution $N(0,\sigma^2)$. The
value of $\sigma^2$ is determined by the signal-to-noise ratio
%
\begin{equation}
\mbox{SNR}=\frac{\sigma_{g}^2}{\sigma^2},
\end{equation}
where $\sigma_{g}^2$ is the sample variance of $g(X_i) = \int
X_{i}(t)\beta(t) \,dt$. The simulation results presented in this article
are under $\mbox{SNR}=9$, which is also considered in Example~4 of
\citet{tib96}. For the Lasso method, \citet{Zou06} observed
that smaller SNR usually yields smaller relative prediction error.
For each of the settings, we use $n=100$ training observations to fit
the model. The optimal tuning parameter is selected by using one of the
following methods: (i)~validating by a separate validation (SV) data
set of the same size; (ii) 5-fold cross-validation (CV); (iii) AIC and
(iv) BIC [\citet{Zou07}] given below:
%
\begin{eqnarray}
\mbox{AIC} &=& \frac{\|Y-\hat{g}(X)\|^2}{n\hat{\sigma}^2}+\frac{2}{n}\hat{d}f,
\\
%
\mbox{BIC} &=& \frac{\|Y-\hat{g}(X)\|^2}{n\hat{\sigma}^2}+\frac
{\log(n)}{n}\hat{d}f,
\end{eqnarray}
where $\hat{d}f$ is the number of nonzero elements of $\hat{\eta}$ in
model (\ref{model}). We estimate $\sigma^2$ by the refitted
cross-validation method introduced in \citet{Fanetal2012}. We then
generate $n={}$10,000 test observations to calculate the mean squared
errors (MSEs) of the corresponding selected models. The procedure is
repeated 100 times and the average MSEs and their standard errors (SE)
for each of the models are presented in Table~\ref{tab1}. We also
report the percentages of correctly identified zero regions and nonzero
regions in Table~\ref{tab1}. We can see that all four methods perform
reasonably well, while the nonpractical SV method performs the best.
The CV method seems to have a nice trade-off between sparsity and
prediction accuracy. Averages of the estimates of $\beta(t)$ using the
CV method over 100 replications are shown in Figure~\ref{fig1}.

\begin{table}
\tabcolsep=4pt
\caption{Average MSEs with standard errors (SE, in parentheses) and
average percentage of correctly identified nonzero and zero elements
over 100 replications for 1D cases}\label{tab1}
\begin{tabular*}{\tablewidth}{@{\extracolsep{\fill}}@{}lc cc cccc@{}}
\hline
& & &&\multicolumn{4}{c@{}}{\textbf{Average percentage (\%)}}\\[-6pt]
& & & &\multicolumn{4}{c@{}}{\hrulefill}
\\
& & \multicolumn{2}{c}{\textbf{Average MSE (SE) ($\bolds{\times10^{-3}}$)}}    &  \multicolumn{2}{c}{\textbf{Case 1}} &\multicolumn{2}{c@{}}{\textbf{Case 2}}\\[-6pt]
& & \multicolumn{2}{c}{\hrulefill} & \multicolumn{2}{c}{\hrulefill} &\multicolumn{2}{c@{}}{\hrulefill}
\\
\textbf{Type}  & \textbf{Method}&\multicolumn{1}{c}{\textbf{Case 1}} &\multicolumn{1}{c}{\textbf{Case 2}} & \textbf{Nonzero} & \textbf{Zero} & \textbf{Nonzero} & \textbf{Zero}\\
\hline
$B$-spline &SV&  0.11 (0.05) & 0.19 (0.08) & 84.30 & 69.20 & 96.00 & 57.26\\
&CV&  0.15 (0.11) &  0.23 (0.11) & 82.95 & 69.68 & 95.03 & 58.90\\
&BIC&  0.60 (1.96) &  1.63 (3.10) & 72.70 & 96.14 & 83.26 & 79.36\\
&AIC&  0.56 (1.96) &  1.56 (3.12) & 75.80 & 93.80 & 82.51 & 82.27
\\[3pt]
Fourier &SV&  0.65 (0.30) & 1.20 (0.49) &84.00 & 70.59 &95.87 & 58.93\\
&CV&  0.92 (0.56) &  1.46 (0.63) & 82.30 & 71.39 & 95.56 & 55.76\\
&BIC& 1.12 (0.86) &  10.62 (20.69) &72.75 & 96.59 & 84.03 & 67.07\\
&AIC& 1.05 (1.28) &  10.29 (20.82) &75.80 & 93.64& 83.85 & 69.01\\
\hline
\end{tabular*}
\end{table}

\begin{figure}

\includegraphics{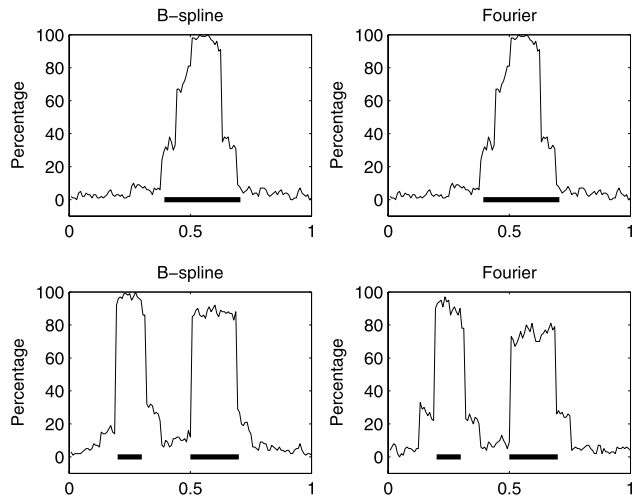}

\caption{Frequency of rejecting the null hypothesis $\beta(t)=0$ using
5-fold cross-validation based on 100 permutation repetitions. The thick
solid horizontal segments indicate the true nonzero regions. The top
panel is for case~1, and the bottom panel is for case~2.}
\label{fig2}
\end{figure}

We also conduct permutation tests to assess the significance of the
regularized estimates of $\beta(t)$. For each of the training data
sets, we generate 200 permutation data sets by randomly shuffling the
response values. Using the same model selection technique for each of
the 200 permutation data sets, 200 sets of $\hat{\beta}_{\mathrm{perm}}(t)$
are obtained. At each $t_{j}$, $j=1,\ldots,p$, the two-sided critical
values are set to be the 2.5th and 97.5th percentiles of $\hat{\beta
}_{\mathrm{perm}}(t_{j})$ for the significance level of 0.05. Supposing the
null hypothesis is $\beta(t_{j})=0$ at each $t_{j}$, we will reject the
null hypothesis if $\hat{\beta}(t_{j})$ is within the critical region.
Repeating this permutation process 100 times, we can compute the
percentages that we reject the null hypothesis at each $t_{j}$. The
results of the permutation tests using the CV method are presented in
Figure~\ref{fig2}, which shows high rejection frequency in the regions
where $\beta(t)$ is nonzero.

\subsection{3D simulation}\label{sec3.2}
\begin{figure}

\includegraphics{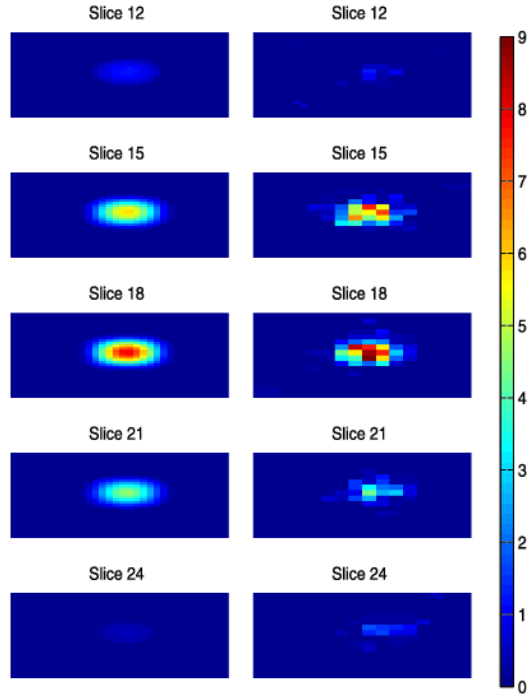}

\caption{The left panel is true $\beta(u,v,w)$ at five selected slices
and the right panel is the average of $\hat{\beta}(u,v,w)$ estimated
using 5-fold cross-validation over 100 replications at the same five slices.}
\label{fig3}
\end{figure}

For the 3D case, we generate the following type of images $ X(u,v,w) =
X^{*}(u,v,w) + {\mathcal E}(u,v,w)$ with
\begin{eqnarray*}
&& X^{*}(u,v,w)
\\
&&\qquad = a_{0}+a_{1}\sin(2\pi
u)+a_{2}\cos(2\pi u)+a_{3}\sin(2\pi v)
\\
&&\quad\qquad{}+a_{4}\cos(2\pi v)+a_{5}\sin(2\pi w)+a_{6}
\cos(2\pi w),\qquad 0\leq u,v,w\leq1,
\end{eqnarray*}
where $a_{i}\sim N(0,1)$ and ${\mathcal E}(u,v,w)\sim N(0,\sigma
_{\mathcal E}^2)$ with $\sigma_{\mathcal E}^2$ similarly defined as in
the 1D case. For simplicity, we record $32\times32\times32$ equally
spaced measurements in the unit cube.
We define the coefficient function $\beta(u,v,w)$ as follows:
\begin{eqnarray*}
&& \beta(u,v,w)
\\
&&\qquad  = \cases{ a \bigl(\sin(bu+c)+1 \bigr) \bigl(\sin(bv+c)+1
\bigr)
\bigl( \sin(bw+c)+1 \bigr),
\vspace*{3pt}\cr
\hspace*{11pt}\qquad \mbox{if }(u-7\pi/40)^2+(v-7
\pi/40)^2+(w-7 \pi/40)^2 \leq(3\pi/40)^2;
\vspace*{3pt}\cr
0,\qquad \mbox{otherwise},}
\end{eqnarray*}
where $a=1/8$, $b=40/3$ and $c=\pi/6$. Note that $\beta(u,v,w)$ is zero
outside a ball that is located in the center of the unit cube. The
error term $\varepsilon$ in model~(\ref{3dmodel}) also follows a normal
distribution $N(0,\sigma^2)$ with $\mbox{SNR} = 9$. We generate 400
training images and apply the 3D Haar wavelet transform to decompose
each image and obtain the wavelet coefficient matrix. Optimal tuning
parameters are selected using the same procedures as for the 1D case.
The results are summarized in Table~\ref{tab2}. Figure~\ref{fig3}
illustrates the comparison of the true $\beta(u,v,w)$ and the mean
estimates of $\beta(u,v,w)$ over 100 replications at five different
slices, which shows that our approach performs reasonably well in
detecting signals based on visual inspection and on the high percentage
of correctly identified nonzeros and zeros reported in Table~\ref{tab2}.

\begin{table}[b]
\tabcolsep=2.6pt
\tablewidth=260pt
\caption{Average MSEs with standard errors (SE, in parentheses), and
average percentages of correctly identified nonzero and zero elements
over 100 replications for 3D case}\label{tab2}
\begin{tabular*}{\tablewidth}{@{\extracolsep{\fill}}@{}lcccc @{}}
\hline
 &  & \multicolumn{2}{c@{}}{\textbf{Average percentage (\%)}}\\[-6pt]
 & & \multicolumn{2}{c@{}}{\hrulefill}\\
\textbf{Method} &\textbf{Average MSE (SE) ($\bolds{\times10^{-4}}$)} & \textbf{Nonzero} & \textbf{Zero}\\
\hline
SV & 0.97 (0.29)& 77.15 & 61.97\\
CV & 1.21 (0.51)& 74.25 & 57.04\\
BIC & 4.78 (1.52)& 39.48 & 99.42\\
AIC & 4.11 (2.13)& 41.86 & 98.74\\
\hline
\end{tabular*}
\end{table}


\section{ADNI PET analysis}\label{sec4}

The FDG PET data used in the preparation of this article were obtained
from the ADNI database (\href{http://adni.loni.ucla.edu}{adni.loni.ucla.edu}). The ADNI was launched in
2003 by NIA, NIBIB, FDA, private pharmaceutical companies and nonprofit
organizations, as a \$60 million, 5-year public-private partnership.
The primary goal of ADNI has been to test whether serial magnetic
resonance imaging (MRI), PET, other biological markers, and clinical
and neuropsychological assessment can be combined to measure the
progression of MCI and early AD. Determination of sensitive and
specific markers of very early AD
progression is intended to aid researchers and clinicians in developing
new treatments, monitoring treatment effectiveness, and lessening the
time and cost of clinical trials. The Principal Investigator of this
initiative is Michael W. Weiner, MD, VA Medical Center and University
of California, San Francisco. ADNI is the result of efforts of many
co-investigators from a broad range of academic institutions and
private corporations, and subjects have been recruited from over 50
sites across the U.S. and Canada. The initial goal of ADNI was to
recruit 800 adults, ages 55 to 90, to participate in the research,
approximately 200 cognitively normal older individuals to be followed
for 3 years, 400 people with MCI to be followed for 3 years and 200
people with early AD to be followed for 2 years. For up-to-date
information, see \href{http://www.adni-info.org}{www.adni-info.org}.

\begin{table}
\tabcolsep=0pt
\caption{Demographics of ADNI participants ($n=403$)}\label{tab3}
\begin{tabular*}{\tablewidth}{@{\extracolsep{\fill}}lccc@{}}
\hline
\textbf{Category} &\textbf{Sex (\% male)} & \textbf{Age (SD)}& \textbf{MMSE (SD)}\\
\hline
NC ($n=102$)&$60.8\%$&80.9 (4.7)&28.9 (1.1)\\
MCI ($n=206$)&$67.0\%$&79.7 (7.3)&27.2 (1.7)\\
AD ($n=95$)&$58.9\%$&80.4 (7.5)&23.4 (2.1)\\
\hline
\end{tabular*}
\end{table}

\begin{figure}

\includegraphics{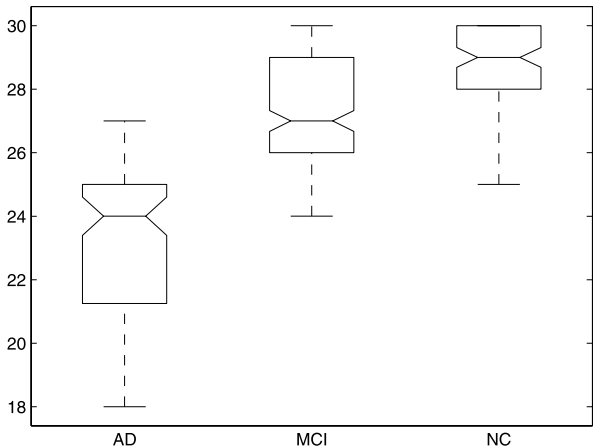}

\caption{Box plots of MMSE scores among AD, MCI and NC.}
\label{fig4}
\end{figure}

In the ADNI's FDG PET study, the injected dose of FDG was \mbox{$5.0\pm0.5$}
mCi, and subjects were scanned from 30 to 60 minutes post-injection
acquiring 6 five-minute frames. The scans were preprocessed by the
following steps: each frame was co-registered to the first frame of the
raw image file; six co-registered frames were averaged to create a
single 30-minute PET image; each subject's co-registered, averaged PET
image from the baseline PET scan was reoriented into a standard
$160\times160\times96$ voxel image grid with 1.5 mm cubic voxels and
the anterior-posterior axis of the subject is parallel to a line
connecting the anterior and posterior commissures (the AC--PC line). It
should be noted that the number of voxels in each image is over 2.4
million, so the approach via linear programming, as in \citet
{Jamesetal2009}, is too computationally expensive for this application.
The data set consists of 403 scans, including 102 NCs, 206 subjects
with MCI and 95 subjects diagnosed with AD. The demographic
characteristics of the 403 subjects are described in Table~\ref{tab3}.
The goal of our analysis is to identify brain subregions that are most
closely related to MMSE scores; we therefore choose not to adjust for
age and other demographic variables. The summary of MMSE scores among
the three groups of participants is given in Figure~\ref{fig4}.
We treat each PET image as a realization of the 3D functional predictor
and then fit the 3D functional linear regression model (\ref{3dmodel}).
The voxel values outside the brain are set to zero prior to
implementing the 3D Haar wavelet transform. We further reduce the
computational cost by excluding those columns of the wavelet
coefficient matrix where all the elements are zero.

In terms of applying the 3D Haar wavelet transforms to each subject's
PET image data, we consider all the possible levels of the Haar wavelet
decompositions. Two tuning parameters are therefore included in the
model selection procedure: the level of the 3D Haar wavelet
decomposition and the lasso regularization parameter.

First, we employ a 10-fold cross-validation to evaluate the predictive
power of the proposed method. Specifically, for each set of 10\%
observations, we leave them out as a test set, use the remaining data
as the training data to fit a model (including selecting the tuning
parameters via 5-fold cross-validation) and compute the prediction
error on the data points that have been left out. We aggregate these
quantities by using the predictive R-square given by $1 - \sum(y_i -
\hat{y}_{i,-i})^2 / \sum(y_i - \bar{y})^2$, where $ \hat{y}_{i,-i}$
denotes the predicted value of $y_i$ calculated by using the estimator
obtained from the training data generated from the cross-validation.
The result is 0.26 for the ADNI data set, whereas the standard R-square
is 0.51,
suggesting a moderate predictive power of the model.

\begin{figure}

\includegraphics{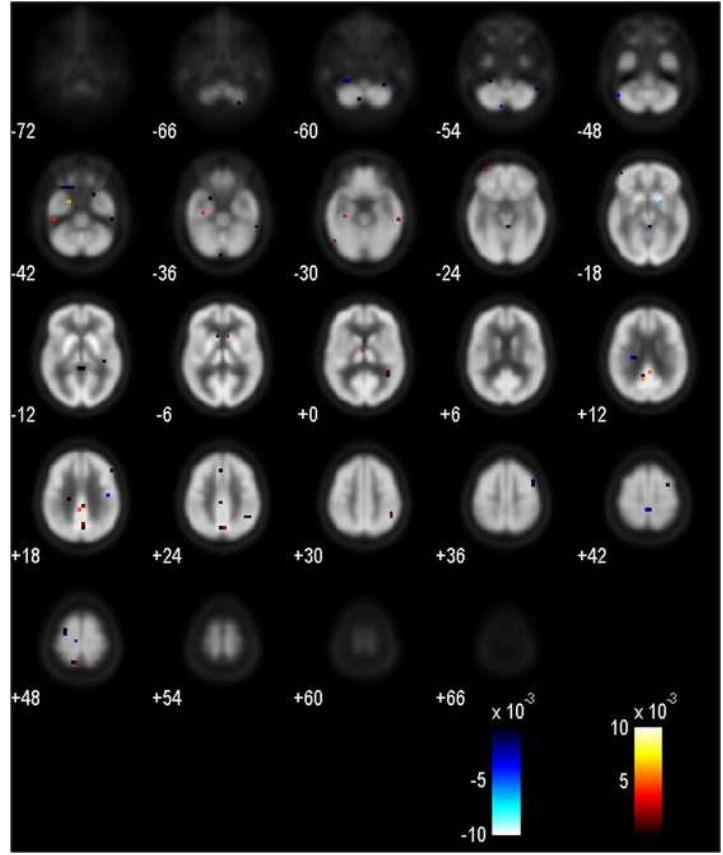}

\caption{Clusters of voxels identified using our approach for the ADNI data.}
\label{fig5}
\end{figure}


\begin{figure}

\includegraphics{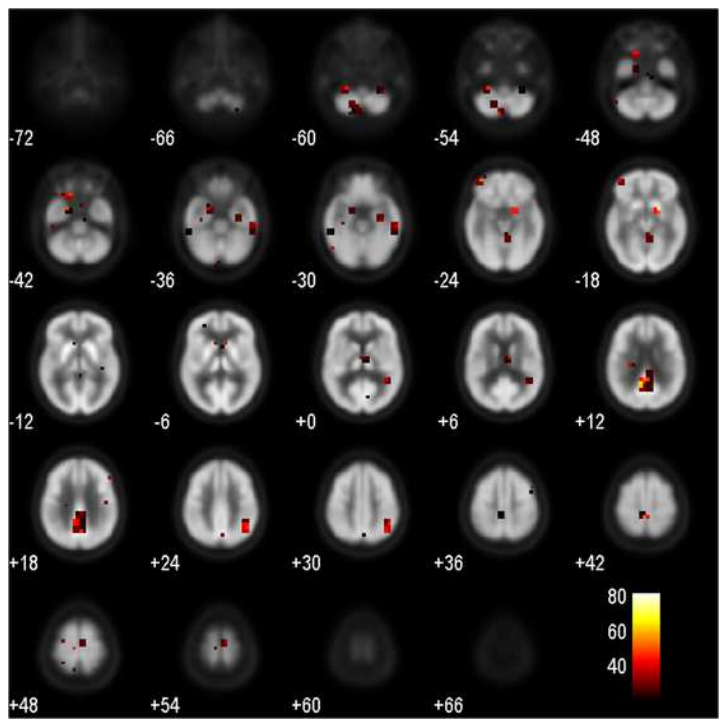}

\caption{Bootstrap inclusion frequencies of the voxels over 100
bootstrap samples.}
\label{fig6}
\end{figure}

\begin{figure}[b]

\includegraphics{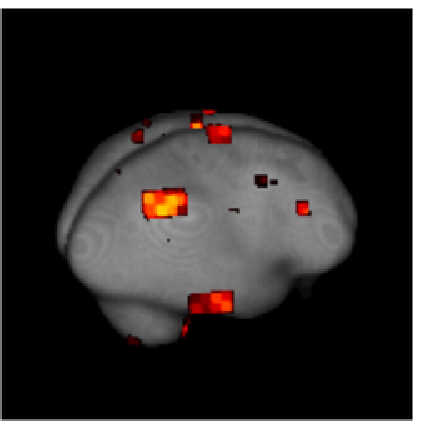}

\caption{Locations of the frequently selected voxels in the 3D sagittal view.}
\label{fig7}
\end{figure}

Second, we investigate the voxels that are selected by our method. We
use 5-fold cross-validation to the full data set to choose the optimal
set of tuning parameters. The identified clusters of voxels [$\hat{\beta
}(u,v,w)\ne0$] are shown on selected axial slices in Figure~\ref
{fig5}, which are presented from the bottom of the brain to the top.
The clusters of voxels with hot colors show a positive association to
prediction of MMSE scores, whereas those with cold colors show a
negative association. Each small square represents a small cluster of voxels.
To assess the significance of the selected voxels, similar to what we
have done in simulation studies, we permute the response variable MMSE
score 200 times. It turns out that 95.3\% of the selected voxels are
significant at the 5\% level. In addition to this pointwise testing, we
also consider the global test described by \citet{Nicholsetal01},
which provides a way to control the family-wise error rate by comparing
$\hat{\beta}(t_{j})$ to a ``maximal statistic.'' It turns out that only
15.6\% of the selected voxels are significant at the 5\% level, which
is more conservative than the pointwise testing procedure.
To further evaluate the stability of the selection, we generate 100
bootstrap samples and for each bootstrap sample, we apply our method
including the tuning parameter selection via 5-fold cross-validation.
Similar approaches also have been employed by other researchers, such
as \citet{Sauerbrei92}, \citet{Royston08} and \citet
{Meinshausen10}.
To summarize the results, we count the number of times that each voxel
is selected over 100 bootstrap samples and denote it as the bootstrap
inclusion frequency (BIF). The voxel BIFs are presented in Figure~\ref
{fig6}. The locations of these more frequently selected voxels are also
presented in the 3D sagittal view in Figure~\ref{fig7} for ease of
understanding. It can be seen that the highly selected brain regions
agree well with the results in Figure~\ref{fig5}.
We note that the clusters of voxels identified in our analysis shown in
Figures~\ref{fig5} and \ref{fig6} reveal high associations of the
expected anatomical regions with cognitive deficits. For example, the
orange ones on slices ``$+$12'' and ``$+$18'' in Figure~\ref{fig5} and
the big cluster on the same slices in Figure~\ref{fig6} indicate that
the posterior cingulate/precuneus cortex is significantly related to
cognitive impairment; the blue ones on slices ``$-$60,'' ``$-$54'' and
``$-$48'' in Figure~\ref{fig5} and the clusters on the same slices in
Figure~\ref{fig6} suggest that the medial temporal/hippocampal cortex
is also closely involved; the red ones on slices ``$-$42,'' ``$-$36'' and
``$-$30'' in Figure~\ref{fig5} and the corresponding clusters on the same
slices in Figure~\ref{fig6} correspond to the lateral temporal cortex.
Many studies have demonstrated that the most prominent metabolic
abnormalities are found in these regions; see, for example, \citet
{Fosteretal84}, \citeauthor{Minoshimaetal95} (\citeyear{Minoshimaetal95,Minoshimaetal97}), \citet{Muelleretal05}. In our study, we have
particularly found the most predictive voxels of the cognitive
impairment in these regions. Other involved regions include the
superior lateral parietal cortex and the frontal cortex, which are all
known to be related to the progression of Alzheimer's disease.

\section{Discussion}\label{sec5}
In this article we propose a highly effective Haar wavelet-based
regularization approach that can be easily applied to analyzing
multidimensional functional data. Analysis of the PET image data
demonstrates that our \mbox{approach} is useful in finding brain subregions
that are most responsible for cognitive impairment in elderly people.
It has great potential to efficiently assist the diagnosis of disease
in neuroimaging studies, yielding easily interpretable results. Our
approach is also computationally fast because of the implementation of
the coordinate descent algorithm with the MATLAB glmnet package. For
example, the real data analysis of 403 subjects' PET image data can be
finished in less than two hours on a 64-bit Intel Xeon 3.33~GHz server
with about 35~GB of RAM, including the selection of tuning parameters.
We should note that another practical advantage of our approach is that
the wavelet transform itself can reduce the dimensionality of the large
volume of brain image data. As a result, we can then apply the proposed
approach on reduced data sets. In such situations, although the
resolution of the original PET images is decreased, the results remain
largely the same since the related subregions are usually not comprised
of a single voxel but of a cluster of voxels.

\section*{Acknowledgments}
Data used in preparation of this article were obtained from the
Alzheimer's Disease Neuroimaging Initiative (ADNI) database\break
(\href{http://adni.loni.ucla.edu}{adni.loni.ucla.edu}). As such, the investigators within the ADNI
contributed to the design and implementation of ADNI and/or provided
data but did not participate in analysis or writing of this report. A
complete listing of ADNI investigators can be found at
\url{http://adni.loni.usc.edu/wp-content/uploads/how\_to\_apply/ADNI\_Acknowledgement\_List.pdf}.

Data collection and sharing for this project
were funded by the Alzheimer's Disease Neuroimaging Initiative (ADNI)
(National Institutes of Health Grant U01 AG024904). ADNI is funded by
the National Institute on Aging, the National Institute of Biomedical
Imaging and Bioengineering, and through generous contributions from the
following: Abbott; Alzheimer's Association; Alzheimer's Drug Discovery
Foundation; Amorfix Life Sciences Ltd.; AstraZeneca; Bayer HealthCare;
BioClinica, Inc.; Biogen Idec Inc.; Bristol-Myers Squibb Company; Eisai
Inc.; Elan Pharmaceuticals Inc.; Eli Lilly and Company; F. Hoffmann-La
Roche Ltd and its affiliated company Genentech, Inc.; GE Healthcare;
Innogenetics, N.V.; IXICO Ltd.; Janssen Alzheimer Immunotherapy
Research \& Development, LLC.; Johnson \& Johnson Pharmaceutical
Research \& Development LLC.; Medpace, Inc.; Merck \& Co., Inc.; Meso
Scale Diagnostics, LLC.; Novartis
Pharmaceuticals Corporation; Pfizer Inc.; Servier; Synarc Inc.; and
Takeda Pharmaceutical Company. The Canadian Institutes of Health
Research is providing funds to support ADNI clinical sites in Canada.
Private sector contributions are facilitated by the Foundation for the
National Institutes of Health (\href{http://www.fnih.org}{www.fnih.org}). The grantee organization
is the Northern California Institute for Research and Education, and
the study is coordinated by the Alzheimer's Disease Cooperative Study
at the University of California, San Diego. ADNI
data are disseminated by the Laboratory for Neuro Imaging at the
University of California, Los Angeles.

The authors would like to thank Professor Karen Kafadar, the Associate
Editor and two referees for their helpful and constructive comments.


\begin{supplement}[id=suppA]
\stitle{Appendix} 
\slink[doi]{10.1214/14-AOAS736SUPP} 
\sdatatype{.pdf}
\sfilename{aoas736\_supp.pdf}
\sdescription{The online supplementary material contains the technical
appendix showing the theoretical results of the proposed approach and
an illustrative example showing the desirable feature of Haar wavelets.}
\end{supplement}

%

\printaddresses

\end{document}